\documentclass[journal,1p]{elsarticle}
\usepackage{amsmath}
\usepackage{amssymb}
\usepackage{amsfonts}
\usepackage{graphicx}
\usepackage{fullpage}
\usepackage{graphics}
\usepackage{multirow}
\usepackage{epsfig}
\usepackage{epstopdf}

\usepackage{newfloat}
\usepackage{subcaption}
\usepackage{gensymb}
\usepackage[table]{xcolor}

\usepackage{varioref}
\usepackage{hyperref}
\usepackage{bm}

\usepackage{lineno,hyperref}
\modulolinenumbers[5]

\newcolumntype{H}{>{\setbox0=\hbox\bgroup}c<{\egroup}@{}} % Hiding entire column of a table without removing it.

\graphicspath{{./}{ER/main/}{ER/SM/}{ER/}{SF/}{./images/}}

\journal{Journal of \LaTeX\ Templates}

\begin{document}

\begin{frontmatter}

\title{Principal eigenvector localization and centrality in networks: revisited}
% \tnotetext[mytitlenote]{Fully documented templates are available in the elsarticle package on \href{http://www.ctan.org/tex-archive/macros/latex/contrib/elsarticle}{CTAN}.}

\author[mymainaddress]{Priodyuti Pradhan}
% \address{Radarweg 29, Amsterdam}
% \fntext[myfootnote]{Since 1880.}
\author[mymainaddress]{Angeliya C. U.}

% \author[mymainaddress,mysecondaryaddress]{}
% \ead[url]{http://iiti.ac.in/people/~sarika/}

\author[mymainaddress,mysecondaryaddress]{Sarika Jalan\corref{mycorrespondingauthor}}
\cortext[mycorrespondingauthor]{Corresponding author}
\ead{sarikajalan9@gmail.com}

\address[mymainaddress]{Complex Systems Lab, Discipline of Physics, Indian Institute of Technology Indore, Khandwa Road, Simrol, Indore-453552, India}
\address[mysecondaryaddress]{Discipline of Biosciences and Biomedical Engineering, Indian Institute of Technology Indore, Khandwa Road, Simrol, Indore-453552, India}

\begin{abstract}
Complex networks or graphs provide a powerful framework to understand importance of individuals and their interactions in real-world complex systems. Several graph theoretical measures have been introduced to access importance of the individual in systems represented by networks. Particularly, eigenvector centrality (EC) measure has been very popular due to its ability in measuring importance of the nodes based on not only number of interactions they acquire but also particular structural positions they have in the networks. Furthermore, the presence of certain structural features, such as the existence of high degree nodes in a network is recognized to induce localization transition of the principal eigenvector (PEV) of the network's adjacency matrix. Localization of PEV has been shown to cause difficulties in assigning centrality weights to the nodes based on the EC. We revisit PEV localization and its relation with failure of EC problem, and by using simple model networks demonstrate that in addition to the localization of the PEV, the delocalization of PEV may also create difficulties for using EC as a measure to rank the nodes. Our investigation while providing fundamental insight to the relation between PEV localization and centrality of nodes in networks, suggests that for the networks having delocalized PEVs, it is better to use degree centrality measure to rank the nodes.
\end{abstract}

\begin{keyword}
\texttt{Complex networks \sep Eigenvector centrality\sep Eigenvector localization\sep Network Centrality}
% \MSC[2010] 00-01\sep  99-00
\end{keyword}

\end{frontmatter}

% \linenumbers

\section{Introduction}
Networks provide a mathematical framework to model interactions among individual entities and to analyze their combined behavior \cite{rev_Strogatz_2001, rev_Jalan_2017}. A network is composed of the nodes representing the individual units of a complex system, and edges representing the interactions among these units. For instance, in a transportation system, cities represent the nodes and the routes among them represent the links or edges. Social media like Facebook can also be seen as edges of friendship where people are the nodes. It is often important to have the information of ``most influential'' or ``central nodes'' in a network. For example, assuming a population of a city as a social network where we want to spread news, spreading of information will be faster if we pass the news to the central person or a group of people having more connections. To understand the relevance of a node in a network, different types of centrality measures have been proposed \cite{newman2010, pevec_nat_phys_2013}. Particularly, measures based on degree centrality, betweenness centrality, closeness centrality, and EC, have been shown to be successful in assigning centrality weights to the nodes in a network. The degree centrality identifies a node as a central node based on the number of edges connected to it, whereas, EC measures devised by Bonacich is based on how important its neighboring nodes are and is calculated by taking into account the weights of the neighboring nodes as well \cite{evec_centrality_2007}. Further, the EC value for each node in a network can easily be calculated from the PEV entry of the corresponding adjacency matrix \cite{newman2010}. In other words, the EC vector of a network corresponds to the PEV of the network's adjacency matrix. 

Despite considerable success of EC in ranking the nodes of a network \cite{evec_cen_global_bird_2017,evec_cen_brain_2010,human_brain_centrality_2017,physica_eigenvec_cent_2016, physica_eigenvec_cent_2018,network_controlability_2017,earthquake_multiplex_2018,hypergraph_centrality_2018}, using the concepts of random matrix theory, Martin {\it et al.} analytically demonstrated that under certain circumstances PEV might undergo a localization transition where most of the weights get concentrated on a few nodes leading to a failure of EC \cite{loc_centrality_2014, evec_localization_2013}. Note that PEV is said to be localized when irrespective of the network size few entries of the vector take large constant weights while rest of the entries receive tiny weights. There exists another extreme case for PEV, i.e., the delocalized state. PEV is said to be delocalized when all the entries in PEV receive almost the same weight irrespective of the network size \cite{loc_centrality_2014, satorras_localization2016}. Importance of eigenvector localization has been realized in an array of fields and problems. For examples, localization of eigenvectors has been investigated in quantum physics \cite{anderson_loc_1985,localization_in_mat_2016}, mathematics \cite{loc_math_1_2010, loc_math_2_2013, loc_math_3_2003}, designing of approximation algorithm  \cite{approx_algo_2015,machine_learning_loc_2016}, numerical linear algebra, matrix gluing, structural engineering, computational quantum chemistry  \cite{loc_invariant_subspace_2011, anderson_loc_linear_alg_1999}, and in data science \cite{faster_least_square_2011}. Particularly, investigation of PEV localization corresponds to the steady-state behavior of many linear dynamical models, which includes epidemic spreading, RNA neutral networks, rumor spreading, brain network dynamical models \cite{evec_localization_2017, pevec_nat_phys_2013, hierarchical_2017}. A dynamical system represented by a network having a localized PEV indicates that a few nodes contribute more in the dynamical process and the rest of them have less contribution irrespective of the system size. Similarly, for a delocalized PEV, all the nodes have the same amount of contribution to a corresponding dynamical system.

Using random matrix theory, Ref. \cite{loc_centrality_2014} proposed a structural relationship ($d>c(c+1)$) between the average degree ($c$) and maximum degree ($d$) of a network to observe the localization of PEV, thereafter imposing severe problems to the EC measure. These studies concentrated on finding constraints for a localized PEV and its impact on the EC measures. However, it is not clear what impact a delocalized PEV state has on the EC measure. Additionally, if there exists a relationship between the network's parameters governing its structure, ensuring a delocalized PEV?
 
Using a recently developed wheel-random-regular (WRR) model network \cite{evec_localization_2018}, current study shows 
that not only PEV localization can lead to a problem for the EC in assigning weights to the nodes causing to the failure of EC, but PEV delocalization can also create problems to the EC measure. Furthermore, it is known that for a connected regular graph, PEV is delocalized (Theorem 6 \cite{miegham_book2011}), and therefore the degree centrality and EC provide the same information \cite{evec_centrality_2007}. While it is obvious that random regular networks have delocalized PEV as all the nodes carry the same information in the network, investigations of the WRR model reveal that graphs consisting of heterogeneous degrees can also have delocalized PEVs. Further, WRR model is shown to hold a particular relationship between the largest eigenvalue of its subgraph components \cite{evec_localization_2018}. 
Here, in this paper, using the WRR model, we show that along with the occurrence of localization state, occurrence of a delocalization of PEV can also affect weights assignment to the higher degree nodes based on EC, thereby creating difficulties in accessing relative importance of the nodes, ergo causing the failure of EC.

The article is organized as follows: Section 2 describes the notations and definitions of the mathematical terms. Section 3 illustrates the results demonstrating PEV localization-delocalization for wheel-random-regular, star-random-regular, friendship-random-regular, and scalefree-random-regular networks. It also contains a subsection which discusses the failure of EC measure due to the localization-delocalization transition of PEV. Finally, section 4 summarizes our work and discusses open problems for further investigations.

\section{Methods}
A graph can be represented as $\mathcal{G}=(V, E)$ where $V$ is the set of nodes and $E$ is the set of interactions (links) among them. We denote $|V|=n$ as the number of nodes and $|E|=m$ being the number of edges of $\mathcal{G}$. Here, we consider undirected, unweighted, and connected networks. Hence, the corresponding adjacency matrix can be denoted as ${\bf A}$ and represented easily as 
\begin{equation}\label{eq1}
a_{ij}=
   \begin{cases}
     1       & \quad \text{if nodes $i$ and $j$ are connected} \\
     0  & \quad \text{ Otherwise}\\
   \end{cases}
\nonumber   
\end{equation}
The number of edges to a particular node is referred as its degree denoted as $k_i = \sum_{j=1}^{n} a_{ij}$. The average degree of the network is denoted by $\langle k \rangle$ = $\frac{1}{n}\sum _{i=1}^{n} {k}_{i}$. We refer the maximum degree node or the hub node of $\mathcal{G}$ as $d = \max_{1\leq i \leq n} k_i$.

Here, ${\bf A}$ is a symmetric matrix and consequently has a set of real eigenvalues $\{\lambda_1, \lambda_2,\ldots, \lambda_n\}$. The corresponding orthonormal set of eigenvectors are $\{\bm {v}_1,\bm{v}_2,\ldots,\bm{v}_n\}$ where $$\bm{v}_i=((v_i)_1,(v_i)_2,\ldots,(v_i)_n)^{T}$$ for $i=1,2,\ldots,n$. Further, ${\bf A}$ is a non-negative and irreducible matrix and it follows from the Perron-Frobenius theorem that there exists a positive and simple eigenvalue $\lambda_1$ \cite{linear_algebra}. The eigenvector corresponding to $\lambda_1$ is a unique positive eigenvector ($\bm{v}_1$) referred as the principal eigenvector. 

The EC vector, $\bm{x}=(x_1,x_2,\ldots,x_n)^{T}$ denotes the centrality of all the nodes and $x_i$ can be calculated as
\begin{equation}\label{evec_cent}
x_i = \lambda_1^{-1}\sum_{j=1}^{n} a_{ij} x_j
\end{equation}
It is well known that $\bm{x}$ corresponds to $\bm{v}_1$ \cite{newman2010}. 
Further, we use the inverse participation ratio (IPR) to measure the extent of the PEV localization \cite{loc_centrality_2014, castellano_localization_2017}. This measure had been introduced to quantify the participation of atoms in a normal mode and is similar to the fourth moments in statistics \cite{ipr_1_1980, ipr_2_1970}. The IPR of the PEV can be calculated \cite{loc_centrality_2014, satorras_localization2016,castellano_localization_2017,Goltsev_prl2012} as follows:
\begin{equation}\label{ipr}
Y_{\bm{v}_1} = {\sum}_{j=1}^{n} (v_1)_j^4 
\end{equation}
where $(v_1)_j$ is the $j^{th}$ component of $\bm{v}_1$. A completely localized PEV with components $\bm{v}_{1}=(1,0,\ldots,0)^{T}$ yields an IPR value,  $Y_{\bm{v}_{1}} = 1$, whereas a completely delocalized PEV with component $\bm{v}_{1}=(\frac{1}{\sqrt{n}},\frac{1}{\sqrt{n}},\ldots,\frac{1}{\sqrt{n}})^{T}$ has $Y_{\bm{v}_{1}}=\frac{1}{n}$. In general, PEV is said to be localized if $Y_{\bm{v}_{1}} = \mathcal{O}(1)$ as $n \rightarrow \infty$ and referred to as delocalized if $Y_{\bm{v}_{1}} \rightarrow 0$ as $n \rightarrow \infty$ \cite{Goltsev_prl2012}. 

It is known that for any connected regular graph (every node having the same degree), $\bm{v}_{1}=(\frac{1}{\sqrt{n}},\frac{1}{\sqrt{n}},\ldots,\frac{1}{\sqrt{n}})^{T}$ (Theorem 6 \cite{miegham_book2011}) and thus, $Y_{\bm{v}_{1}}=\frac{1}{n}$. Therefore, for a regular network IPR value of PEV provides the lower bound. Hence, a sparse as well as a dense regular network both will have a delocalized PEV. Next, if we consider a disconnected graph where each node is isolated without having any interaction with any one and having a self-loop, adjacency matrix will be nothing but an identity matrix and for which we can choose $\bm{v}_{1}=(1,0,\ldots,0)^{T}$ leading to $Y_{\bm{v}_{1}} = 1$. However, if we consider a connected network having non-negative entries, all the entries of the PEV is positive (from Perron-Frobenius theorem). Hence, IPR value of the PEV should be in the range $1/n \leq Y_{\bm{v}_{1}} < 1$ for $n \geq 2$. However, to test whether the PEV is localized or not for IPR being in the range $1/n < Y_{\bm{v}_1} < 1$, we adopt the procedure proposed for the detection of the Anderson localization \cite{ipr_11_1972} and which was recently used to measure the eigenvector localization in complex networks \cite{loc_centrality_2014, satorras_localization2016, castellano_localization_2017}. According to this procedure, one should calculate the IPR value of PEV for different network sizes. If $Y_{\bm{v}_1}$ tends to have a constant value as $n \rightarrow \infty$, PEV is localized, otherwise it is delocalized \cite{castellano_localization_2017}.

\begin{figure}
\centering
\includegraphics[width=0.45\linewidth]{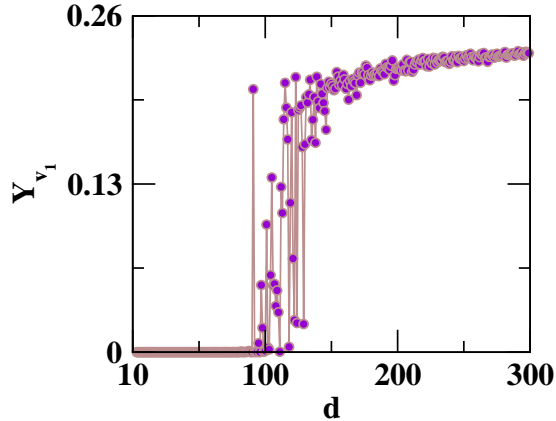}
\caption{Localization transition in the PEV for hub node size being larger than the square of the average degree of the networks ($d > c(c + 1)$) \cite{loc_centrality_2014}. For $d$ being much larger, IPR value shows a drastical increase leading to a non-zero constant value. We consider a connected random graph of size $n=100000$ and average degree $c=10$, where $n^{th}$ node acts as a hub node and degree of the $n^{th}$ node varies from $10$ to $300$.} 
\label{loc_tran}
\end{figure}

\begin{figure*}[t]
\centering
\includegraphics[width=5in, height=3in]{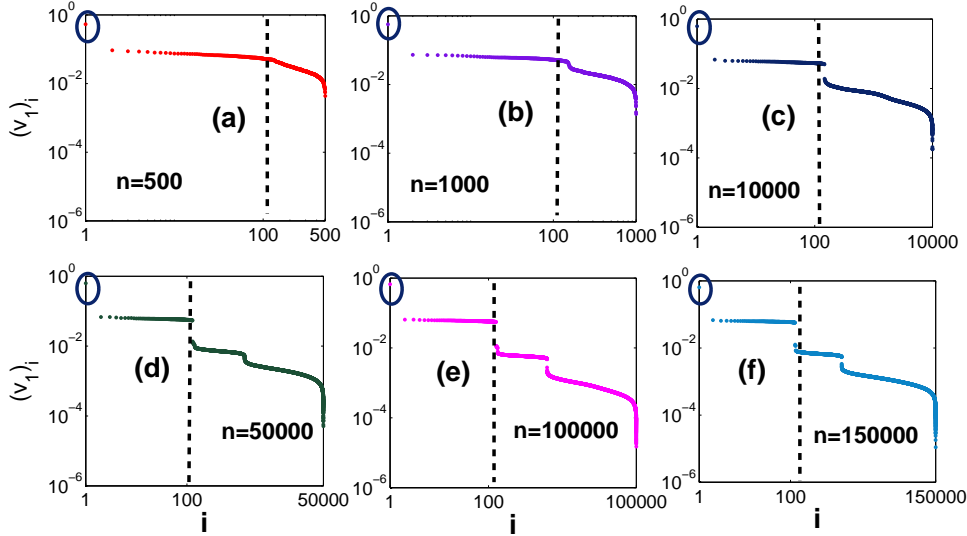}
\caption{Sorted localized PEV entries ($(v_1)_i$ for $i=1,2,\ldots,n$) for different network size. The average degree ($c=10$) and maximum degree ($d=130$) remains fixed and which satisfies $d>c(c + 1)$. The PEV entries corresponding to the hub node (marked with a circle) and its adjacent nodes receive large constant weights as $n \to \infty$. The PEV entries corresponding to the nodes which are not connected to the hub node (after dotted vertical lines) gradually show a decrease, eventually becoming close to zero as $n \to \infty$.}
\label{pev_behavior}
\end{figure*}
\section{Results and Discussion} 
We first consider a graph model which has only one hub node \cite{loc_centrality_2014, evec_localization_2013}. This model network has localized PEV. The PEV entries experience minor changes with the change in the network size confirming its localization. Furthermore, using a 
wheel-random-regular (WRR) model networks \cite{evec_localization_2018} and its variants, we demonstrate that non-regular networks in addition of having a localized PEV can also show delocalized PEV \cite{evec_localization_2018}. Further, with the help of the WRR model network, we show that the delocalization of PEV can also cause failure to the EC as it does not assign sufficiently large weights to the higher degree nodes. All the data and codes used in this paper are available at GitHub repository \cite{codes_data_evec_centrality}.

\subsection{Localization transition in Random Graph Model}
In the random graph (RG) model \cite{loc_centrality_2014}, a random subgraph ($\mathcal{G}_{RG}$) of size $n-1$ is generated with the connection probability between a pair of vertices being $p= c/(n-1)$ for $n \rightarrow \infty$. The random subgraph is generated using the algorithm in \cite{fast_gnp_model}. Further, $n^{th}$ node is included in $\mathcal{G}_{RG}$ such that it connects to the $n-1$ existing nodes of the random subgraph with a probability $d/(n-1)$, where $d>>c$. Hence, the expected number of connections to the $n^{th}$ node should be $d$, thereby $n^{th}$ node will be the hub node. Further, using random matrix theory, it has been shown in Ref. \cite{loc_centrality_2014} that if the size of the hub node is larger than $c(c+1)$, localization transition occurs. Further, the PEV entry corresponding to the hub node and its immediate neighbors are expected to have constant values, and which should only depend on $c$ and $d$. However, rest of the nodes which are not adjacent to the hub node receive a vanishing weight as $n \rightarrow \infty$. Note that $c$ is the average degree of the random subgraph containing $n-1$ number of nodes.

Fig. \ref{loc_tran} indicates that for a hub node of size $d$, $d<c(c + 1)$, IPR value of PEV is small and in fact it lies close to zero.
However, as $d$ becomes greater than $c(c+1)$, there arises a sudden jump in the IPR value (Fig. \ref{loc_tran}) as also illustrated in \cite{loc_centrality_2014}. Though, it is evident from Fig. \ref{loc_tran}, that for fix values of $n$ and $c$, if we form a hub node of size $d$ such that it is larger than  $c(c+1)$, IPR value of PEV is large. However, it is not clear that with an increase in the network size by fixing $c$ and $d$, whether IPR value remains fixed to a large value, and how exactly the PEV entries behave?

Figure \ref{pev_behavior} plots sorted PEV entries ($(v_1)_i$) for different network sizes by fixing $c$ and $d$ such that $d>c(c+1)$. For each value of $n$, the PEV entry corresponding to the hub node adopts a large value (marked with a circle in Fig. \ref{pev_behavior}), and successive PEV entries become approximately equal to each other forming a horizontal band (Fig. \ref{pev_behavior}). Additionally, one can notice that PEV entries in the horizontal band correspond to those nodes which are directly connected to the hub node. After the horizontal band, the PEV entries show a gradually decrease and become close to the zero as $n \to \infty$ (Fig. \ref{pev_behavior}). Hence, in the limit of large $n$, the size of the hub and its neighboring nodes play a vital role in the occurrence of localization transition of PEV. Note that we choose $c$ such that the random subgraph is always connected even for large values of $n$. In the following, we use a few other simple models to demonstrate the localization-delocalization transition as a consequence of single edge rewiring of PEV and relation of this transition with the behavior of EC measure.

\begin{figure}[t]
\centering
\includegraphics[width=3.1in, height=1.3in]{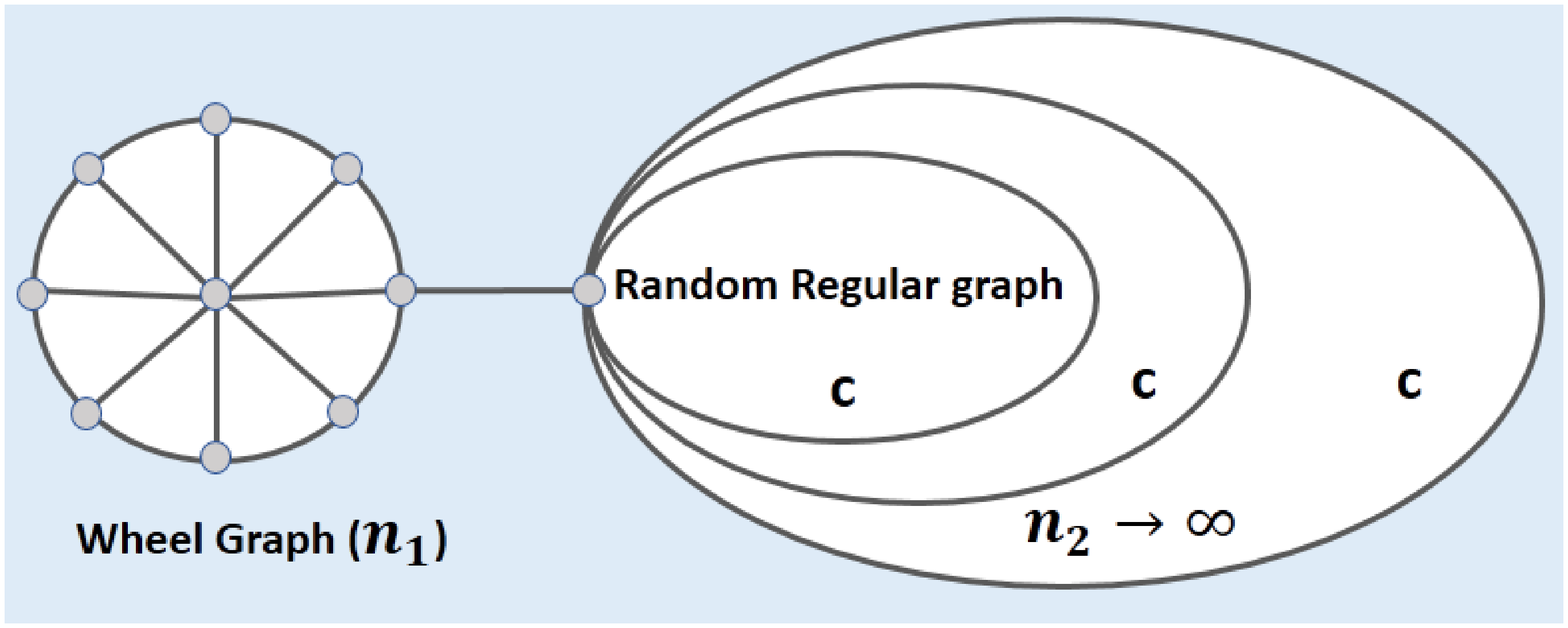}
\caption{Schematic representation of the wheel-random-regular model networks ($\mathcal{G}_{WRR}$). Here, $n_1$ is the number of nodes in the wheel graph and $n_2$ is the number of nodes in the random regular network having an average degree of $c$. With an increase in the number of nodes in the random regular network while fixing $c$, the hub node size of wheel graph remains unchanged.}
\label{schematic}
\end{figure}

\subsection{Wheel-Random-Regular Model}
Recently, a wheel-random-regular (WRR) model was developed in Ref. \cite{evec_localization_2018} which provides a simple method, instead of the random matrix theory, to derive a condition to form a model network having the highly localized PEV. Further, the rewiring of a few special sets of edges in this model network is shown to lead the delocalization transition of PEV. This section demonstrates that occurrence of the localization-delocalization transition of PEV in WRR model networks as a consequence of single edge rewiring creates difficulties in weights assignment to the nodes based on EC, thereby leading to the failure of EC. 

The WRR model consists of a random regular graph and a wheel graph. Let us denote it as $\mathcal{G}_{WRR}$ (Fig. \ref{schematic}). This model network manifests both the localization as well as the delocalization of PEV, occurrence of which is decided by the largest eigenvalue relation of the individual graph components \cite{evec_localization_2018}. We denote the wheel graph as $\mathcal{W}=\{V_\mathcal{W},E_\mathcal{W}\}$ where $|V_\mathcal{W}|=n_1$ is the number of nodes and $|E_\mathcal{W}|=2(n_1-1)$ is the number of edges in $\mathcal{W}$. Further, the random regular graph is denoted as $\mathcal{R}=\{V_\mathcal{R},E_\mathcal{R}\}$ where $|V_\mathcal{R}|=n_2$ is the number of nodes and $|E_\mathcal{R}|=\frac{n_2c}{2}$ is the number of edges with each node having degree $3 \leq c \leq n_{2}-2$. We generate the random regular graph using the algorithm in \cite{random_regular_graph}. Further, it is known that for a wheel and a random regular graph, the largest eigenvalues are as follows \cite{eigval_wheel_graph, graph_spectra_rowlinson}
\begin{equation}\label{eig_vals}
 \lambda_1^{\mathcal{W}} = \sqrt{n_1}+1\; \text{and}\; \lambda^\mathcal{R}_{1}=c 
\end{equation}
We combine a wheel graph and a random regular graph such that 
\begin{equation}\label{eval_relation}
\begin{split}
\text{(a)}\; \lambda_1^{\mathcal{W}} & > \lambda_1^{\mathcal{R}}\; \text{or}\;\; \text{(b)}\;
\lambda_1^{\mathcal{W}} =\lambda_1^{\mathcal{R}}+\epsilon\;\text{where} \;  0<\epsilon<1
\end{split}
\end{equation}
and as learned from Ref. \cite{evec_localization_2018} this leads to occurrence of highly localized PEV. To construct $\mathcal{G}_{WRR}$ by holding the relation in Eq. (\ref{eval_relation}) requires the network parameters ($c$, $n_1$ and $n_2$) of $\mathcal{W}$ and $\mathcal{R}$. One can observe that substitute Eq. \ref{eig_vals} in Eq. \ref{eval_relation}, one can easily find the size of $\mathcal{W}$ as follows
\begin{equation} \label{vertices_wheel_relation}
\text{(a)}\;n_1>(c-1)^2\; \text{or}\;\text{(b)}\; n_1=\lceil (c-1+\epsilon)^2 \rceil
\end{equation}
where ($\lceil \rceil$) is the ceiling function. For the random regular graph, we can choose any arbitrary size ($n_2$) and average degree ($c$) such that $c n_2$ is even. Hence, for a particular value of $c$, Eq. (\ref{vertices_wheel_relation}) implicitly ensures the validity of Eq. \ref{eval_relation}. In $\mathcal{G}_{WRR}$ all the nodes corresponding to $\mathcal{R}$ component has degree $c$ except one node having degree $c+1$ which connects to $\mathcal{W}$ component. Simultaneously, all the peripheral nodes of $\mathcal{W}$ has degree $3$ except one node of degree $4$ which connects to $\mathcal{R}$ component. From Eq. (\ref{vertices_wheel_relation}), the maximum degree node of the $\mathcal{W}$ component in $\mathcal{G}_{WRR}$ has degree 
\begin{equation}\label{hub_node_wheel}
\text{(a)}\; d  > (c-1)^2 -1 \;\text{or}\;\text{(b)}\; \;d  = \lceil (c-1+\epsilon)^2 \rceil -1,\; \text{where}\; d=n_1-1\\
\end{equation}
and which is always be greater than $c$ ($3 \leq c\leq n_2-1$). Therefore, Eq. (\ref{hub_node_wheel}), establishes a structural relationship between the size of the hub node ($d$) and average degree ($c$) of random regular graph, which can also be holds true for WRR model networks having the localized PEV. Hence, Eq. (\ref{hub_node_wheel}) ensures that the maximum degree node of $\mathcal{G}_{WRR}$ will always come from the wheel graph component of $\mathcal{G}_{WRR}$.
\begin{figure}[t]
\centering
\includegraphics[width=5.3in, height=1.5in]{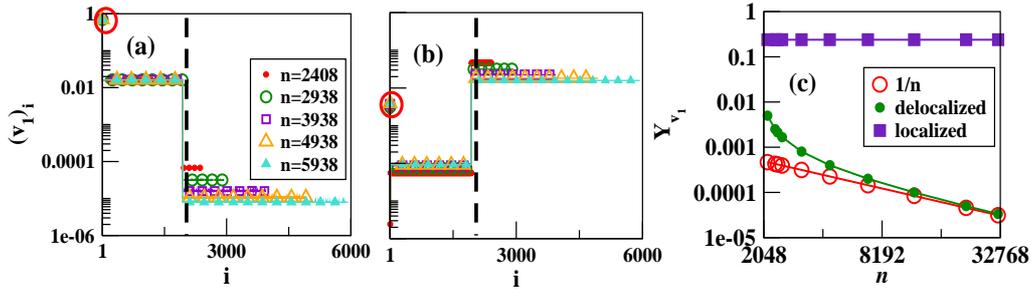}
\caption{Value of localized PEV entries of (a) network constructed by combining a wheel graph ($\mathcal{W}$) with a random regular ($\mathcal{R}$) graph. We choose $n_1=1938$, $n_2 = 470$ and $c=45$ satisfying $\lambda_1^{\mathcal{W}} =\lambda_1^{\mathcal{R}}+\epsilon$, $\epsilon=0.00002$ and yielding a network ($\mathcal{G}_{WRR}$) with $n=n_1+n_2=2408$ nodes and $m=14806$ edges. For $\mathcal{G}_{WRR}$, IPR value of PEV is equal to $0.2389$. Next, by fixing size of $\mathcal{W}$ ($n_1=1938$), we increase size of $\mathcal{R}$ by keeping $c$ constant. This arrangement leads to $\lambda_1^{\mathcal{W}} =\lambda_1^{\mathcal{R}}+\epsilon$ and keeps the PEV entries same for the hub as well as its adjacent nodes as $n \rightarrow \infty$. (b) The same network but by removing an edge connected to the hub node in $\mathcal{W}$ and adding it between a pair of nodes in $\mathcal{R}$. This rewiring yields delocalization transition in PEV. By increasing the size of $\mathcal{G}_{WRR}$ by including nodes to $\mathcal{R}$ keeping the $c$ fixed, and the eigenvalue relation $\lambda_1^{\mathcal{W}} <\lambda_1^{\mathcal{R}}$ holds true, and PEV entries take very less values for the hub and its neighboring nodes as $n \rightarrow \infty$. (c) IPR values of PEV ($Y_{\bm{v}_{1}}$) for the wheel-random-regular (WRR) graph as a function of combined network size $n$. For each value of $n$, we consider two WRR graphs, first WRR graph is generated by holding the relation in Eq. (6) and PEV is localized ($\blacksquare$). The second WRR graph has an edge rewired, and PEV becomes delocalized ($\bullet$). For reference to a delocalization, we  plot $1/n$ ($\bigcirc$) as a function of $n$.}
\label{pev_deloc_large}
\end{figure} 

In the following discussions, we primarily focus on the localization-delocalization transition as a consequence of single edge rewiring and for which we consider $0<\epsilon<1$. Importantly, upon changing the bound to $\epsilon>0$, we get part (a) of Eqs. (4), (5) and (6). In Appendix B, we analytically show that by holding $\lambda_1^{\mathcal{W}} > \lambda_1^{\mathcal{R}}$, PEV entries of $\mathcal{G}_{WRR}$ corresponding to the wheel subgraph contribute more to IPR as compared to that of the regular graph part and vice-versa.

\subsubsection{Localization transition in WRR model}
First, we perform an experiment to show that $\mathcal{G}_{WRR}$ contains a localized PEV. We construct $\mathcal{G}_{WRR}$ by combining a wheel graph and a random regular graph by satisfying Eq. (\ref{eval_relation}) and which gives a large IPR value. Next, to ensure localization of PEV, we fix the average degree of $\mathcal{R}$ and increase the number of nodes in $\mathcal{R}$, resulting in an increase in the size of $\mathcal{G}_{WRR}$. In the other words, increasing $n_2$ leads to an increase in the number of nodes ($n$) in $\mathcal{G}_{WRR}$, however, the network keeps satisfying $\lambda_1^{\mathcal{W}} =\lambda_1^{\mathcal{R}}+\epsilon$. We observe that PEV entries remain almost constant for the hub and its adjacent nodes (Fig. \ref{pev_deloc_large}(a)), as well as IPR remains fixed to a constant value (Fig. \ref{pev_deloc_large}(c)) indicating localization of PEV.

\subsubsection{Delocalization transition in WRR model}
Next, to witness the delocalized state, we consider the same model as the above and simply rewire an edge from the $\mathcal{W}$ component to the $\mathcal{R}$ component of $\mathcal{G}_{WRR}$. We denote the modified graph as $\widetilde{\mathcal{G}}_{WRR}$, and the imperfect wheel and random regular as $\widetilde{\mathcal{W}}$ and $\widetilde{\mathcal{R}}$, respectively. 
In $\mathcal{G}_{WRR}$ (Fig. \ref{schematic}), just by removing an edge connected to the hub node of $\mathcal{W}$ and adding it between a pair of nodes in $\mathcal{R}$ yields $\widetilde{\mathcal{G}}_{WRR}$ with $\lambda_1^{\widetilde{\mathcal{W}}} =\lambda_1^{\mathcal{W}}-\delta = \lambda_1^{\mathcal{R}}+\epsilon-\delta$ such that $\delta>\epsilon$.  
Upon performing one such rewiring, there is a drastic change in the PEV entries (Fig. \ref{pev_deloc_large}(b)) leading to a small IPR value indicating delocalization. However, to ensure this delocalization in PEV upon such rewiring, we increase the size of $\widetilde{\mathcal{G}}_{WRR}$ by including more number of nodes to the random regular subgraph without changing the average degree $c$ (Fig. \ref{schematic}) leading to an unchange of the eigenvalue relation for both of the components ($\lambda_1^{\widetilde{\mathcal{W}}} =\lambda_1^{\mathcal{R}}+\epsilon-\delta$). The IPR shows a value close to $1/n$ as $n \rightarrow \infty$ which confirms delocalization of PEV (Fig. \ref{pev_deloc_large}(c)). 
Interestingly, this network has a hub node of the significant size ($d=\lceil (c-1+\epsilon) \rceil-2$), however, due to an occurrence of the delocalization in PEV, EC is unable to assign high weights to the hub and its neighboring nodes resulting in a failure of EC measure (Fig. \ref{pev_deloc_large}(b)). Further, one can say that there exists a single point transition for localization-delocalization of PEV.
Notably, in the delocalized state, the PEV weights flip between $\widetilde{\mathcal{W}}$ and $\widetilde{\mathcal{R}}$ of $\widetilde{\mathcal{G}}_{WRR}$, as well as PEV entry weight corresponding to the hub node becomes tiny (Fig. \ref{pev_deloc_large}(b)). To conclude, we have found at least one network structure where the delocalization of PEV creates a problem to EC. Note that Ref. \cite{evec_localization_2018} illustrates that for such situations, there exists a localized second largest eigenvector.

\begin{figure}[t]
\centering
\includegraphics[width=4in, height=3.2in]{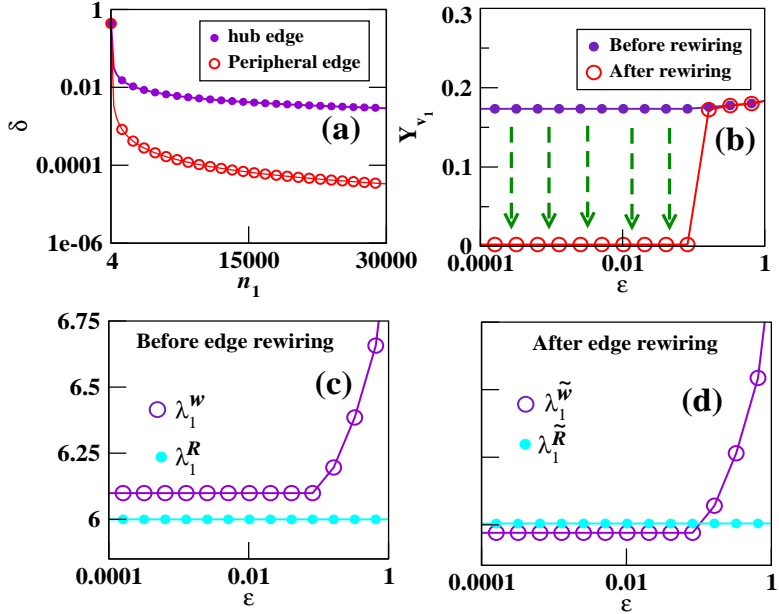}
\caption{ (a) Measure the difference between $\delta = \lambda_1^{\mathcal{W}}-\lambda_1^{\widetilde{\mathcal{W}}}$ as a function of wheel graph size ($n_1$) after removing hub edge ($\bullet$) and peripheral edge ($\circ$). IPR value of the combined network and the largest eigenvalue of the individual component as a function of $\epsilon$ to demonstrate that a particular value of $\epsilon$ is required to witness sudden change in the IPR value as a consequence of a single edge rewiring. (b) IPR of localized and delocalized graph and (c) plots $\lambda_1^{\mathcal{W}}$ \& $\lambda_1^{\mathcal{R}}$ and (d) plots $\lambda_1^{\widetilde{\mathcal{W}}}$ \& $\lambda_1^{\widetilde{\mathcal{R}}}$. The size of the wheel graph is calculated from Eq. (\ref{vertices_wheel_relation}), where parameters of the random regular graph is $n_2=500$ and $c=6$.} 
\label{eps_vs_ipr}
\end{figure}
Next let us explain that moving one edge from $\mathcal{W}$ to $\mathcal{R}$ in $\mathcal{G}_{WRR}$ involves two steps (i) removing an edge from $\mathcal{W}$ component and (b) including it in the $\mathcal{R}$ component. We know that removal of an edge from a connected graph always leads to a  decrease in the largest eigenvalue, whereas addition of a new edge in a network leads to an increase the largest eigenvalue (Proposition 1.3.10, \cite{graph_spectra_rowlinson}). To track the amount of decrement ($\delta$) in $\lambda_1^{\mathcal{W}}$ as a consequence of single edge removal, we perform numerical simulations for different values of $n_1$. It is clear from Fig. \ref{eps_vs_ipr}(a) that removal of an edge connected to the hub node of $\mathcal{W}$ leads to a significant decrease in $\lambda_1^{\mathcal{W}}$ whereas removal of an edge connected to the peripheral nodes in $\mathcal{W}$ leads to negligible change in $\lambda_1^{\mathcal{W}}$. Therefore, to capture the sudden change in the PEV localization upon a single edge rewiring, we focus on those edges of $\mathcal{W}$ which are connected to the hub node. Further, upon addition the edge to the $\mathcal{R}$ leads to an increment in $\lambda_1^{\mathcal{R}}$ which is negligible and we assume $\lambda_1^{\widetilde{\mathcal{R}}}=\lambda_1^{\mathcal{R}}$. It indicates that before and after a single edge rewiring, $\lambda_1^{\mathcal{W}}$ is affected substantially. 
As we have already mentioned that for the localized PEV, $\lambda_1^{\mathcal{W}}=\lambda_1^{\mathcal{R}}+\epsilon$, $0<\epsilon<1$ whereas for a delocalized PEV, $\lambda_1^{\widetilde{\mathcal{W}}}=\lambda_1^{\mathcal{R}}+\epsilon-\delta$ provided $\delta>\epsilon$ upon a sinlge edge rewiring. Hence, single edge removal leads to a decrement in $\lambda_1^{\mathcal{W}}$ such that $\delta$ is greater than $\epsilon$ leading to delocalization transition. Fig. \ref{eps_vs_ipr}(a) convey that $\delta$ itself is a small quantity for large $n_1$. Thus, if we consider $\epsilon<<1$, $\delta$ can easily take value greater than $\epsilon$ upon a single edge rewiring leading to the delocalization transition. However, if we consider $\epsilon<1$ (Fig. \ref{eps_vs_ipr}(b)) or $\epsilon>1$, we have to either remove more number of edges (nodes) from the $\mathcal{W}$ or increase the average degree of $\mathcal{R}$ to adjust the eigenvalue relation between the individual component to witness the delocalization transition. Fig. \ref{eps_vs_ipr}(c) and (d) depict the behavior of $\lambda_1^{\mathcal{W}}$ and $\lambda_1^{\mathcal{R}}$ as $\epsilon$ value changes. The figures portray that before the rewiring eigenvalue relation follows as $\lambda_1^{\mathcal{W}}>\lambda_1^{\mathcal{R}}$, and after the rewiring eigenvalue relation changes to $\lambda_1^{\widetilde{\mathcal{W}}}<\lambda_1^{\widetilde{\mathcal{R}}}$.

Further, we show that instead of WRR model, one can observe the single localization-delocalization transition point for the other models as well. The only condition which holds good should be that one component ($\mathcal{C}_1$) should have a hub node and another ($\mathcal{C}_2$) should be random regular such that they satisfy the eigenvalue relation ($\lambda_1^{\mathcal{C}_1}>\lambda_1^{\mathcal{C}_2}$). In the following, we perform the investigation by replacing $\mathcal{W}$ with a star, friendship \cite{friendship_graph_2013} and scalefree \cite{newman2010} graph, and show that regulating the largest eigenvalues of the subgraph components, one can observe the localization-delocalization transition for the combined network.

\subsubsection{Localization transition in other graph models}
\begin{figure}[t]
\centering
\includegraphics[width=5in, height=1.4in]{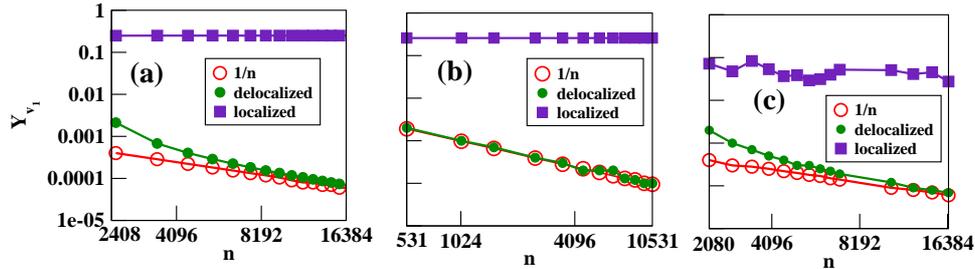}
\caption{(a) IPR values of PEV ($Y_{\bm{v}_{1}}$) for the star-random-regular (SRR) graph as a function of combined network size $n$. For each value of $n$, we consider two SRR graphs, first SRR graph is generated by holding $\lambda_{1}^{\mathcal{S}}>\lambda_{1}^{\mathcal{R}}$ and PEV is localized ($\blacksquare$). The second SRR graph has delocalized PEV by holding $\lambda_{1}^{\mathcal{S}}<\lambda_{1}^{\mathcal{R}}$ ($\bullet$). For reference to a delocalization, we plot $1/n$ ($\bigcirc$) as a function of $n$. (b) Same as (a) but for the friendship-random-regular (FRR) networks; (c) Same as (a)  but for the scalefree-random-regular (SFRR) networks.}
\label{loc_deloc_tran}
\end{figure}

As demonstrated for the WRR model, we first show that there exists a localized state of PEV for the star random regular model ($\mathcal{G}_{SRR}$). We consider star graph $\mathcal{S}=\{V_\mathcal{S},E_\mathcal{S}\}$ having $|V_\mathcal{S}|=n_{1}$ number of nodes, $|E_\mathcal{S}|=n_{1}-1$ number of edges and  $\lambda_1^{\mathcal{S}} = \sqrt{n_{1}-1}$ \cite{graph_spectra_rowlinson}. $\mathcal{G}_{SRR}$ is constructed by combining a $\mathcal{S}$ and a $\mathcal{R}$ which satisfies $\lambda_1^{\mathcal{S}} > \lambda_1^{\mathcal{R}}$ and leads to a large IPR value. Next, to ensure the localization of PEV in $\mathcal{G}_{SRR}$, we fix the average degree of $\mathcal{R}$ and increase the number of nodes in $\mathcal{R}$ resulting in an increase in the size of $\mathcal{G}_{SRR}$. However, the network keeps satisfying $\lambda_1^{\mathcal{S}} > \lambda_1^{\mathcal{R}}$. We observe that IPR remains fixed to a constant value as $n$ changes (Fig. \ref{loc_deloc_tran}(a)) indicating localization of PEV.

Finally, we replace $\mathcal{W}$ in the WRR model with a friendship ($\mathcal{F}$) and a scalefree network ($\mathcal{SF}$), respectively. We find that again by holding the eigenvalue relation between the two subgraph component, PEV can be made localized (Fig. \ref{loc_deloc_tran}(b) and (c)). Table 1 summarizes the network parameters to construct WRR, SRR, FRR and SFRR model networks having localized PEV. 

\subsubsection{Delocalization transition in other graph models}
In $\mathcal{G}_{SRR}$, removing an edge connected to the hub node and adding it between a pair of nodes in $\mathcal{R}$ yields $\lambda_1^{\mathcal{S}} < \lambda_1^{\mathcal{R}}$ where $\lambda_1^{\mathcal{S}}$ and $\lambda_1^{\mathcal{R}}$ are eigenvalues of $\mathcal{S} $ and $\mathcal{R}$, respectively. Upon performing one such rewiring, there exists a drastic change in the PEV entries yielding a small IPR value indicating delocalization of PEV. To ensure the occurrence of the delocalization transition in PEV, we increase the size of $\mathcal{G}_{SRR}$ by including more number of nodes to $\mathcal{R}$ without changing the average degree $c$ as in the WRR model. The eigenvalue relation for both of the components keep holding true ($\lambda_1^{\mathcal{S}} <\lambda_1^{\mathcal{R}}$) for $n \rightarrow \infty$, and the IPR value comes closer to $1/n$ which confirms the PEV delocalization (Fig. \ref{loc_deloc_tran}(b)). 
\begin{table}[b]
\centering
\begin{tabular}{|c|c|c|c|c|}
\hline
\multirow{2}{*}{}$\mathcal{G}$&WRR&SRR&FRR&SFRR \\  
\hline
$n_1$&$n_1>(c-1)^2$& $n_1>c^2+1$ &$n_1>\lceil(c^2-\frac{1}{2})^2+\frac{3}{4}\rceil$&$n_1$ \\
$\lambda_1^{\mathcal{C}_1}$ &$\lambda_1^{\mathcal{W}}=1+\sqrt{n_1}$ &$\lambda_1^{\mathcal{S}}=\sqrt{n_{1}-1}$&$\lambda_1^{\mathcal{F}}=\frac{1}{2}+\frac{1}{2}\sqrt{4n_1-3}$ \cite{friendship_graph_2013}&$\lambda_1^{SF} \approx \max \{\sqrt{d},\frac{\langle k^2 \rangle}{\langle k \rangle}\}$ \cite{castellano_localization_2017}\\
$\lambda_1^{\mathcal{C}_2}$ &$\lambda_1^{\mathcal{R}}=c$, $c\geq 3$ &$\lambda_1^{\mathcal{R}}=c$, $c\geq 2$&$\lambda_1^{\mathcal{R}}=c$, $c\geq 2$&$\lambda_1^{\mathcal{R}}=c$, $c\geq 2$\\
$d$&$d  > (c-1)^2 -1$ &$d  > c^2$&$d>\lceil(c^2-\frac{1}{2})^2+\frac{3}{4}\rceil-1$&$d$\\
\hline
\end{tabular}
\caption{Various network parameters of wheel-random-regular (WRR), star-random-regular (SRR), friendship-random-regular (FRR) and scalefree-random regular (SFRR) networks which provides localized PEV state.}
\label{graph_construction}
\end{table}
Similarly, for the friendship-random regular and scalefree-random regular models, one can adjust the eigenvalue relation between individual components to make delocalization of PEV (Fig. \ref{loc_deloc_tran}(b) and (c)).

\section{Failure of EC measure}
\begin{figure}[t]
\centering
\includegraphics[width=3.8in, height=1.6in]{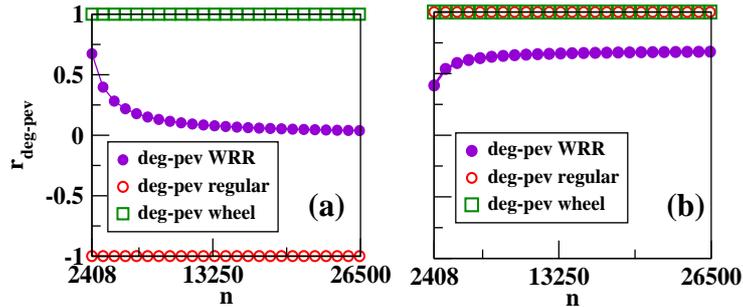}
\caption{ Pearson correlation between normalized degree vector and PEV 
of WRR ($\bullet$), corresponding to random regular component of WRR ($\circ$) and corresponding to wheel component of WRR ($\Box$), respectively
as a function of $n$ for (a) localized PEV state; (b) delocalized PEV state.}
\label{deg_pev_corr}
\end{figure}

The model graph structure demonstrating a peculiar behavior of EC is artificially constructed; however, these investigations, helps in having a better understanding of the effects of PEV localization and delocalization in complex networks. Importantly for all the models (WRR, SRR, FRR, SFRR), there can exist the localized as well as delocalized PEV state is an interesting observation. As EC weights correspond to the PEV entries weights, the localization transition in the PEV is accompanied with assigning almost constant weights to the hub and its neighboring nodes, and very tiny weights to the rest of the nodes in the network. Therefore, it is predetermined that in the localized environment, EC will always assign a large weight to the hub node, followed by comparatively smaller weights to all its neighboring nodes. Rest of the nodes will receive negligible weights though their degrees can be higher than those of the neighboring nodes of the hub node (Fig. \ref{pev_deloc_large}(a)). 
Further, calculation of the Pearson correlation coefficient between the normalized degree vector (in Euclidean norm) and PEV which is denoted by $r_{deg-pev}$ reveals that for the WRR model network, having a localized PEV, degree and PEV are uncorrelated (Fig. \ref{deg_pev_corr}(a)). Importantly,
there exists a negative correlation between those degrees and the PEV entries which correspond to the random regular component of the network, and a positive correlation between the degrees and PEV entries corresponding to the wheel component. These situations arise as the nodes not connected to the hub node are unable to receive centrality weights despite of having large degrees as compared to that of the nodes directly connected to the hub node. Consequently, it leads to the failure of EC measure for the localized PEV. Note that, to avoid the localization effect in the PEV of the adjacency matrices, the PEV of non-backtracking matrices has been shown to be useful in ranking the nodes in networks \cite{satorras_localization2016}.

Fig. \ref{deg_pev_corr}(b) reflects that for the case of delocalized PEV, $r_{deg-pev}$ is high. Further, $r_{deg-pev}=1$ if it is measured by excluding the hub node degree and the corresponding PEV entry weight. It indicates that both the degree vector and PEV is highly correlated in the network having delocalized PEV state and hence EC is unable to recognize the wheel hub node as an important node. Our study illustrates that for such kind of core-periphery network structures (WRR, SRR, FRR, SFRR model) having delocalized PEV, it is better to use degree centrality instead of EC for measuring the centrality of the nodes as also mentioned in Refs. \cite{evec_centrality_2007,centrality_core_periphery_2016}.

\section{Conclusion}
The current study has focused on understanding the relation between EC and the networks having localized-delocalized PEV in the limit of large $n$. By using a wheel-random-regular model network, we have demonstrated that not only the localization transition of PEV can cause difficulties for EC in assigning weights to the nodes, but also the delocalization transition of PEV can also cause a problem to the EC measure. Based on numerical simulations for large size networks, we demonstrate that for PEV being localized, the size of the network imparts minor effects to the PEV entry weights corresponding to the hub and its neighboring nodes. 

As EC weighs are the same as that of PEV entries, any particular behavior of the PEV entries also gets reflected in the ranking of nodes through EC. Therefore, for a localized PEV, it is predetermined that the hub node and its neighboring nodes will receive significant weights with the rest of the nodes receiving negligible weights causing the failure of EC. Similarly, in the delocalized PEV, EC is unable to assign centrality weight to the higher degree node. As a result, EC becomes inefficient and uninformative for measuring of the centrality of the nodes when PEV is delocalized. Hence, for various model networks current study shows EC fails for the network when (i) PEV is highly localized, (ii) PEV is delocalized. For the case of the delocalization transition, it has been suggested to use degree centrality instead of the EC \cite{evec_centrality_2007}. Note that non-backtracking matrices have been used instead of adjacency matrix to avoid localization effect in the PEV. 

This work portrays that EC measures and the PEV localization have two different perspectives in which the former is used to rank the nodes and later stands as a particular phenomenon which in this paper is focused for predicting difficulties associated with the EC measure. The PEV localization of network is confirmed if there exists a particular arrangement of the nodes and edges such that few entries of the PEV take very large values with rest of the entries taking tiny values, and this arrangement should hold good irrespective of the network size. Though, here we have focused only on PEV, the localization transition can occur to any eigenvector and not necessarily to PEV. By considering the size of the hub node and average degree fixed, we may not achieve the localized PEV as $n \rightarrow \infty$. Nevertheless, by satisfying a particular relation between the size of the hub node and the average degree of the network, we may achieve a network which undergoes to the localization transition as demonstrated for the model networks discussed in this paper. 

Though, the WRR model network depicts such a very typical behavior of PEV in the localized-delocalized state, which may be difficult to observe for real-world systems. However, we know that many real-world networks follow power-law degree distributions and thus contain several large degree nodes,  naturally forming imperfect wheel graph (i.e., star, friendship, etc.). 
Our study offers a platform to understand PEV localization behaviors of real-world systems, as well as to relate them with the network's structural properties by providing fundamental insight to localization and delocalization behavior of eigenvectors of networks \cite{evec_centrality_2018,loc_bipartite_2017}. Furthermore, since eigenvectors and eigenvalues provide information \cite{spectrum_review_2018}
for energy controllability and synchronization of complex networks \cite{spectrum_controlling_2015, sync_multilayer_2016}, the investigation carried out here for PEV of adjacency matrix can be extended for finding localization of eigenvector for other matrix representations of networks. For instance, Laplacian \cite{loc_laplacian_matrix, network_dynamics_loc_lap_2018}, Jacobian \cite{jacobian_matrix_2014} and Hessian \cite{hessian_matrix_loc_2009, hessian_matrix_loc_2014} matrices which are closely related with coupled nonlinear dynamical evolution on networks.

\section{Acknowledgement}
SJ acknowledges CSIR, Govt. of India grant (25(0293)/18/EMR-II), and DAE, Govt. of India grant (37(3)/14/11/2018-BRNS/37131) for financial support. PP acknowledges CSIR, Govt. of India grant (09/1022(0070)/2019-EMR-I) for SRF fellowship. PP is indebted to Travis Martin and M. E. J. Newman for useful discussions on the eigenvector localization and centrality measures. Angeliya C. U. acknowledges IIT Indore for hospitality during her summer internship and IASc (Bengaluru), INSA (New Delhi) and NASI (Allahabad) for financial support under the summer fellowship programme. 

\appendix
\section{Results related to eigenvalues}
We restate the proposition related to removal of an edge with maximum eigenvalue behavior from \cite{graph_spectra_rowlinson},\\

\noindent {\bf Proposition 1.3.10} If $\mathcal{G}-uv$ is the graph obtained from a connected graph $\mathcal{G}$ by deleting the edge $uv$, then $\lambda_1^{\mathcal{G}-uv}<\lambda_1^{\mathcal{G}}$.\\

% \noindent {\bf Theorem [Row6]} Let $\mathcal{G}$ be a connected graph with distinct vertices $h$,$i$,$j$,$k$ such that $h$ is adjacent to $i$, $j$ is not adjacent to $k$. Let $\mathcal{G}^{'}$ be the graph obtained from $\mathcal{G}$ by replacing edge $hi$ with $jk$. If $\bm{x}_j\bm{x}_k \geq \bm{x}_h\bm{x}_i$ then $\lambda_1^{\mathcal{G}^{'}}>\lambda_1^{\mathcal{G}}$. If $\bm{x}_j\bm{x}_k < \bm{x}_h\bm{x}_i$ and $\lambda_1^{\mathcal{G}}-\lambda_2^{\mathcal{G}}>\frac{\bm{x}_h^2+\bm{x}_i^2+\bm{x}_j+\bm{x}_k}{2(\bm{x}_h\bm{x}_i-\bm{x}_j\bm{x}_k)}$ then $\lambda_1^{\mathcal{G}^{'}}<\lambda_1^{\mathcal{G}}$.
% \newpage
\section{Wheel-Random-Regular model}
Our aim is to interpret PEV and largest eigenvalue of $\mathcal{G}$ interms of PEV and largest eigenvalue of $\mathcal{W}$ and $\mathcal{R}$ and find relation between them. We connect a wheel network ($\mathcal{W}$) with an edge to a random regular network ($\mathcal{R}$) such that $\lambda_1^{\mathcal{W}}>\lambda_1^{\mathcal{R}}$ holds true. Hence, from Eq. (\ref{vertices_wheel_relation}.a), $n_1>(c-1)^2$, which indicates that for a fixed $c$, we can choose a $n_1$ and we have no restriction on $n_2$. Therefore, we get a combined graph $\mathcal{G}$ and the corresponding adjacency matrix (${\bf A}\in \Re^{n \times n}$ such that $n=n_1+n_2$) as follows 
\begin{equation}
{\bf A}\bm{x}_{1} =\lambda_1^{\mathcal{G}}\bm{x}_{1}  
\end{equation}
\[
\begin{bmatrix} \mathcal{W}_{n_1 \times n_1} & \mathcal{P}_{n_1 \times n_2} \\ \mathcal{P}^{T}_{n_2 \times n_1} & \mathcal{R}_{n_2 \times n_2} \end{bmatrix}  \left[ \begin{array}{c} {\bm{x}_{1}^1}_{n_1 \times 1} \\ {\bm{x}_{1}^2}_{n_2 \times 1} \end{array} \right]
= \lambda_1^{\mathcal{G}} \left[ \begin{array}{c} {\bm{x}_{1}^1}_{n_1 \times 1} \\ {\bm{x}_{1}^2}_{n_2 \times 1} \end{array} \right] 
\] 
where $\mathcal{P}$ matrix contains only single one. Hence, we have
\begin{equation}\label{main1}
\begin{split}
\mathcal{W}\bm{x}_{1}^1+\mathcal{P}\bm{x}_{1}^2 &= \lambda_1^{\mathcal{G}}\bm{x}_{1}^1\\
\mathcal{P}^{T}\bm{x}_{1}^1+\mathcal{R}\bm{x}_{1}^2 &= \lambda_1^{\mathcal{G}}\bm{x}_{1}^2\\
\end{split}
\end{equation}
where $\bm{x}_{1}^1 \in \Re^{n_1}$ is the upper part and $\bm{x}_{1}^2 \in \Re^{n_2}$ is the lower part of PEV ($\bm{x}_{1}\in \Re^n$) of ${\bf A}$. Moreover $\mathcal{W} \in \Re^{n_1 \times n_1}$ and $\mathcal{R} \in \Re^{n_2 \times n_2}$ are real symmetric matrices. Hence, eigenvectors of $\mathcal{W}$, $\{\bm{v}_{1}^{\mathcal{W}},\bm{v}_{2}^{\mathcal{W}},\ldots,\bm{v}_{n_1}^{\mathcal{W}}\}$ are orthonormal and form a basis for the $n_1$ dimensional real vector space. 
Similarly, eigenvectors of $\mathcal{R}$, $\{\bm{v}_{1}^{\mathcal{R}},\bm{v}_{2}^{\mathcal{R}},\ldots,\bm{v}_{n_2}^{\mathcal{R}}\}$ are orthonormal and form a basis for the $n_2$ dimensional real vector space. 
Therefore, we can represent $\bm{x}_{1}^1$ and $\bm{x}_{1}^2$ as a linear combinations of the eigenvectors of $\mathcal{W}$ and $\mathcal{R}$ as follows,
\begin{equation}\label{first}
\begin{split}
\bm{x}_{1}^1 = \sum_{i=1}^{n_1} c_i \bm{v}_{i}^{\mathcal{W}}\;\;\text{and}\;\;
\bm{x}_{1}^2 &= \sum_{i=1}^{n_2} d_i \bm{v}_{i}^{\mathcal{R}}
\end{split}
\end{equation}
where $c_i \in \Re$ and $d_i \in \Re$ are the scalar quantity respectively.
In Eq. (\ref{first}), multiplying both sides by $\bm{v}_{1}^{\mathcal{W}^{T}}$ and $\bm{v}_{1}^{\mathcal{R}^{T}}$ we get,
\begin{equation}\label{proj1}
\begin{split}
 c_1 = \bm{v}_{1}^{\mathcal{W}^{T}}\bm{x}_{1}^1\;\;\text{and}\;\;
 d_1 &= \bm{v}_{1}^{\mathcal{R}^{T}}\bm{x}_{1}^2
 \end{split} 
\end{equation}
Now, substitute Eq. (\ref{first}) in Eq. (\ref{main1}) and we get
\begin{equation}\label{p1}
\begin{split}
\mathcal{W}\sum_{i=1}^{n_1} c_i {\bm{v}_{i}^{\mathcal{W}}}+\mathcal{P}\bm{x}_{1}^2 &= \lambda_1^{\mathcal{G}}\sum_{i=1}^{n_1} c_i {\bm{v}_{i}^{\mathcal{W}}}\\
\text{[replace $\mathcal{W}\bm{v}_i^{\mathcal{W}}$ with $\lambda_i^{\mathcal{W}}\bm{v}_i^{\mathcal{W}}$ and multiply both sides by $\bm{v}_{1}^{\mathcal{W}^{T}}$ we get]}\\
\lambda_1^\mathcal{W} + \frac{1}{c_1} \bm{v}_{1}^{\mathcal{W}^{T}} \mathcal{P}\bm{x}_{1}^2 &= \lambda_1^{\mathcal{G}} \\
\text{[as $||\bm{v}_{i}^{\mathcal{W}}||_2^2=1$ and $\bm{v}_{i}^{\mathcal{W}} \perp \bm{v}_{j}^{\mathcal{W}}$, $\forall i\neq j$]}\\
\end{split}
\end{equation}
Similarly, from the second equation in (\ref{main1}) we get,
\begin{equation}\label{p2}
\begin{split}
 \lambda_1^\mathcal{R} + \frac{1}{d_1} \bm{v}_{1}^{\mathcal{R}^{T}} \mathcal{P}^{T}\bm{x}_{1}^1 &= \lambda_1^{\mathcal{G}} \\
\end{split}
\end{equation}
Hence, from Eqs. (\ref{p1}) and (\ref{p2}) we get,
\begin{equation}\label{main_relation}
\begin{split}
\lambda_1^\mathcal{W} + \frac{1}{c_1} \bm{v}_{1}^{\mathcal{W}^{T}} \mathcal{P}\bm{x}_{1}^2 &= \lambda_1^\mathcal{R} + \frac{1}{d_1} 
\bm{v}_{1}^{\mathcal{R}^{T}} \mathcal{P}^{T}\bm{x}_{1}^1\\
\lambda_1^\mathcal{W} - \lambda_1^\mathcal{R} &= \frac{1}{d_1} \bm{v}_{1}^{\mathcal{R}^{T}} \mathcal{P}^{T}\bm{x}_{1}^1 - \frac{1}{c_1} \bm{v}_{1}^{\mathcal{W}^{T}} \mathcal{P}\bm{x}_{1}^2\\
\end{split} 
\end{equation}
From the above relation we can say that if $\bm{x}_1$ is localized then we can assume that $\bm{x}_{1}^2$ will be a zero vector infers that $\lambda_1^\mathcal{W} > \lambda_1^\mathcal{R}$ as $\frac{1}{d_1} \bm{v}_{1}^{\mathcal{R}^{T}} \mathcal{P}^{T}\bm{x}_{1}^1$ is always a positive terms.
Further, we know 
\begin{equation}
\lambda_1^\mathcal{W} > \lambda_1^\mathcal{R} %> 0 
\end{equation}
We substitute Eq. (\ref{proj1}) in Eq. \ref{main_relation} and get,
\begin{equation}
\begin{split}
\frac{1}{d_1} \bm{v}_{1}^{\mathcal{R}^{T}} \mathcal{P}^{T}\bm{x}_{1}^1 &> \frac{1}{c_1} \bm{v}_{1}^{\mathcal{W}^{T}} \mathcal{P}\bm{x}_{1}^2\\ %&> 0 \\
\frac{c_1}{d_1} &> \frac{\bm{v}_{1}^{\mathcal{W}^{T}} \mathcal{P}\bm{x}_{1}^2}{\bm{v}_{1}^{\mathcal{R}^{T}} \mathcal{P}^{T}\bm{x}_{1}^1}\\
\frac{\bm{v}_{1}^{\mathcal{W}^{T}} \bm{x}_{1}^1}{\bm{v}_{1}^{\mathcal{R}^{T}} \bm{x}_{1}^2} &> \frac{\bm{v}_{1}^{\mathcal{W}^{T}} \mathcal{P}\bm{x}_{1}^2}{\bm{v}_{1}^{\mathcal{R}^{T}} \mathcal{P}^{T}\bm{x}_{1}^1} \\
\frac{\frac{1}{\beta}(x_1^1)_1 + \frac{\alpha}{\beta} \sum_{i=2}^{n_1} (x_1^1)_i}{\frac{1}{\sqrt{n_2}} \sum_{i=1}^{n_2} (x_1^2)_i} &> \frac{\frac{\alpha}{\beta}(x_1^2)_1}{\frac{1}{\sqrt{n_2}}(x_1^1)_{n_1}} \\
\end{split} 
\end{equation}
\begin{equation}\label{relation1}
\begin{split}
\frac{(n_1-1)(x_{1}^{1})_1 (x_{1}^{1})_{n_1} + (\sqrt{n_1}+1)(x_{1}^{1})_{n_1} \sum_{i=2}^{n_1} (x_{1}^{1})_i}{(\sqrt{n_1}+1)(x_{1}^{2})_1 \sum_{i=1}^{n_2} (x_{1}^{2})_i}  &> 1 \\
\end{split} 
\end{equation}
where $\bm{v}_{1}^{\mathcal{W}}=\biggl(\frac{1}{\beta},\frac{\alpha}{\beta},\ldots,\frac{\alpha}{\beta}\biggr)$ such that $\alpha=\frac{\sqrt{n_{1}}+1}{n_{1}-1}$, $\beta=\sqrt{1+\frac{(\sqrt{n_{1}}+1)^2}{n_{1}-1}}$ \cite{eigval_wheel_graph} and $\bm{v}_{1}^{\mathcal{R}}=(\frac{1}{\sqrt{n_2}},\frac{1}{\sqrt{n_2}},\ldots,\frac{1}{\sqrt{n_2}})$, respectively. 

From Eq. (\ref{relation1}) one can say that holding the relation $\lambda_{1}^{\mathcal{W}} > \lambda_{1}^{\mathcal{R}}$, PEV of the combined network for which maximum contribution comes from the wheel graph part. Even if we vary $n_2 \rightarrow \infty$ the above relation holds true due to the Perron Frobenius theorem that all the PEV entries should receive positive quantity.


\section*{References}
\begin{thebibliography}{0}
\bibitem{rev_Strogatz_2001} S. H. Strogatz, Exploring complex networks, {\it Nature 2001}; {\bf 410}:268.

\bibitem{rev_Jalan_2017} S. Jalan, C. Sarkar, Complex networks: an emerging branch of science, {\it Physics news 2017}; {\bf 47}:3.

\bibitem{newman2010} M. E. J. Newman, \emph{Networks: An Introduction}, Oxford University Press 2010. 

\bibitem{pevec_nat_phys_2013} J. Aguirre, D. Papo, J. M. Buld\'{u}, Successful strategies for competing networks, {\it Nat. Phys. 2013}, {\bf 9}:230.

\bibitem{evec_centrality_2007} P. Bonacich, Some unique properties of eigenvector centrality, {\it Social Networks 2007}; {\bf 29}(4):555-564.

\bibitem{evec_cen_global_bird_2017} L. Reino {\it et al.}, Networks of global bird invasion altered by regional trade ban, {\it Sci. Adv. 2017}; {\bf 3}:e1700783.

\bibitem{evec_cen_brain_2010} G. Lohmann {\it et al.}, Eigenvector Centrality Mapping for Analyzing Connectivity Patterns in fMRI Data of the Human Brain, {\it PloSOne 2010}; {\bf 5}(4):e1023.

\bibitem{human_brain_centrality_2017} H. T. Karim {\it et al.}, Intrinsic functional connectivity in late-life depression: trajectories over the course of pharmacotherapy in remitters and non-remitters, {\it Molecular Psychiatry 2017}; {\bf 22}, 450.

\bibitem{physica_eigenvec_cent_2016} F. A. Parand, H. Rahimi, M. Gorzin, Combining fuzzy logic and eigenvector centrality measure in social network analysis, {\it Physica A 2016}; {\bf 459}:24.

\bibitem{physica_eigenvec_cent_2018} R. R. Yin, Q. Guo, J. N. Yang, J. G. Liu, Inter-layer similarity-based eigenvector centrality measures for temporal networks, {\it Physica A 2018}; {\bf 512}:165. 
 
\bibitem{network_controlability_2017} N. Bof, G. Baggio, S. Zampieri, On the Role of Network Centrality in the Controllability of Complex Networks, {\it IEEE Trans. on Control of Network Science 2017}; {\bf 4}:3.

\bibitem{earthquake_multiplex_2018}N. Lotfi, A. H. Darooneh, F. A. Rodrigues, Centrality in earthquake multiplex networks, {\it Chaos 2018}; {\bf 28}:063113.

\bibitem{hypergraph_centrality_2018} A. R. Benson, Three hypergraph eigenvector centrality, {\it SIAM J Math Data SCI 2019}; 1(2), pp. 293-312.

\bibitem{loc_centrality_2014} T. Martin, X. Zhang, M. E. J. Newman, Localization and centrality in networks, {\it Phys. Rev. E 2014}; {\bf 90}:052808.

\bibitem{evec_localization_2013} R. R. Nadakuditi, M. E. J. Newman, Spectra of random graphs with arbitrary expected degrees, {\it Phys. Rev. E 2013};  {\bf 87}:012803.

\bibitem{satorras_localization2016} R. Pastor-Satorras, C. Castellano, Distinct types of eigenvector localization in networks, {Sec. Rep. 2016}; {\bf 6}:18847.

\bibitem{anderson_loc_1985} P. A. Lee and T. V. Ramakrishnam, Disordered electronic systems {\it Rev. Mod. Phys.} \textbf{57}, 287 (1985).

\bibitem{localization_in_mat_2016} M. Benzi, Localization in Matrix Computations: Theory and Applications, Springer International Publishing AG, (2016).

\bibitem{loc_math_1_2010} Y. Dekel, J. R. Lee, N. Linial, Eigenvectors of Random Graphs: Nodal Domains, Random Structures \& Algorithms {\bf 39}, 39 (2011).

\bibitem{loc_math_2_2013} C. Bordenave and A. Guionnet, Localization and delocalization of eigenvectors for heavy-tailed random matrices {\it Probability Theory and Related Fields} \textbf{157}, 885-953 (2013).

\bibitem{loc_math_3_2003} X. Liu, G. Strang, and S. Ott, Localized eigenvectors from widely spaced matrix modifications,  SIAM J. Discrete Math., 
vol. 16, No. 3, pp. 479-498, (2003).

\bibitem{approx_algo_2015} David F. Gleich, Michael W. Mahoney, Using Local Spectral Methods to Robustify Graph-Based Learning Algorithms, KDD’15, Sydney, NSW, Australia, 10-13, (2015).

\bibitem{machine_learning_loc_2016} Pan Zhang, Robust Spectral Detection of Global Structures in the Data by Learning a Regularization, 29th Conference on Neural Information Processing Systems (NIPS), Barcelona, Spain (2016).

\bibitem{loc_invariant_subspace_2011} C. V\"{o}mel, and B. N. Parlett, Detecting localization in an invariant subspace, Society for industrial and Applied Mathematics (SIAM) {\bf 33}, 3447 (2011).

\bibitem{anderson_loc_linear_alg_1999} U. Elsner, V. Mehrmann, F. Milde, R. A. Romer, and M. Schreiber, The Anderson model of localization: A challenge for modern eigenvalue methods, SIAM J. Sci. Comput., vol. 20, No. 6, pp. 2089-2102, (1999).


\bibitem{faster_least_square_2011} Petros Drineas, Michael W. Mahoney, S. Muthukrishnan, Tam\'{a}s Sarl\'{o}s, Faster least squares approximation, Numer. Math. 117:219-24 (2011). 

\bibitem{evec_localization_2017} P. Pradhan, A. Yadav, S. K. Dwivedi, and S. Jalan, an Optimized evolution of networks for principal eigenvector localization, Phys. Rev. E {\bf 96}, 022312 (2017).

\bibitem{hierarchical_2017} A. Safari, P. Moretti, M. A. Mu\~{n}oz, Topological dimension tunes activity patterns in hierarchical modular networks, New. J. Phys {\bf 19}, 113011 (2017). 

\bibitem{evec_localization_2018} P. Pradhan, S. Jalan, Eigenvalue crossing in principal eigenvector localized networks, arXiv:1810.00243v2 2019.

\bibitem{miegham_book2011}P. V. Mieghem, \emph{Graph Spectra for Complex Networks}, Cambridge University Press, 2011.

\bibitem{linear_algebra} C. D. Meyer, Matrix Analysis and Applied Linear Algebra, SIAM, 2000.

\bibitem{castellano_localization_2017} C. Castellano, R. Pastor-Satorras, Topological determinants of complex networks spectral properties: structural and dynamical effects, {\it Phys. Rev. X 2017}; {\bf 7}:041024.


\bibitem{ipr_1_1980} F. Wegner, Inverse Participation Ratio in $2 + \varepsilon$ Dimensions, {\it Z. Physik B 1980}; {\bf 36}:209.

\bibitem{ipr_2_1970} R. J. Bell, P. Dean, Atomic Vibrations in Vitreous Silica, {\it Discuss. Faraday Soc. 1970}; {\bf 50}:55.

\bibitem{Goltsev_prl2012} A. V. Goltsev, S. N. Dorogovtsev, J. G. Oliveira, J. F. F. Mendes, Localization and Spreading of Diseases in Complex Networks, {\it Phys. Rev. Lett. 2012}; {\bf 109}:128702.

\bibitem{ipr_11_1972} J. T. Edwards, D. J. Thouless, Numerical studies of localization in disordered systems, {\it J. Phys. C 1972}; {\bf 5}:807. 

\bibitem{codes_data_evec_centrality} Our codes and data are available at the following link \url{https://github.com/priodyuti/pev_loc_in_evec_centrality}.

\bibitem{fast_gnp_model} V. Batagelg, U. Brandes, Efficient generation of large random networks, {\it Phys. Rev. E 2005}; {\bf 71}:036113. 

\bibitem{random_regular_graph} J. H. Kim, V. H. Vu, Generating random regular graphs, {\it Proceedings of the thirty-fifth ACM symposium on Theory of computing, San Diego, CA, USA, 2003}; pp. 213-222.

\bibitem{eigval_wheel_graph} I. T. Abu-Jeb, The Determinant of the Wheel graph and conjectures by Yong, {\it Missouri J. Math. Sci. 2006}; {\bf 18}(2):142.

\bibitem{graph_spectra_rowlinson} D. Cvetkovi\'{c}, P. Rowlinson, S. Simi\'{c}, {\it An introduction to the theory of graph spectra}, Cambridge University Press; $1^{st}$ ed. 2009.

\bibitem{friendship_graph_2013} A. Abdollahi, S. Janbaz and M. R. Oboudi, Graphs cospectral with a friendship graph or its complement, Transactions on Combinatorics {\bf 2}(4), 37-52 (2013).

\bibitem{centrality_core_periphery_2016} P. Barucca, D. Tantari, F. Lillo, Centrality metrics and localization in core-periphery networks, {\it J. Stat. Mech. 2016}; 023401. 

\bibitem{evec_centrality_2018} K. J. Sharkey, Localization of eigenvector centrality on networks with a cut-vertex, {\it Phys. Rev. E 2019}; {\bf 99}:012315.

\bibitem{loc_bipartite_2017} F. Slanina, Localization in random bipartite graphs: Numerical and empirical study, {\it Phys. Rev. E 2017}; 
{\bf 95}:052149.

\bibitem{spectrum_review_2018} C. Sarkar, S. Jalan, Spectral properties of complex networks, {\it Chaos 2018}; {\bf 28}:102101.

\bibitem{spectrum_controlling_2015} G. Yan, G. Tsekenis, B. Barzel, J. J. Slotine, Y. Y. Liu, A. L. Barab\'{a}si, Spectrum of controlling and observing complex networks, {\it Nat. Phys. 2015}; {\bf 11}:779.

\bibitem{sync_multilayer_2016} C. I. D Genio, J. G. Garde\~{n}es, I. Bonamassa, S. Boccaletti, Synchronization in networks with multiple interaction layers, {\it Science Advances 2016}; {\bf 2}(11):e1601679.

\bibitem{loc_laplacian_matrix} S. Hata and H. Nakao, Localization of Laplacian eigenvectors on random networks, Sec. Rep. {\bf 7}, 1121 (2017). 


\bibitem{network_dynamics_loc_lap_2018} A. Forrow, F. G. Woodhouse, and J. Dunkel, Functional control of network dynamics using designed laplacian, Phys. Rev. X {\bf 8}, 041043 (2018).

\bibitem{jacobian_matrix_2014} Helge Eberhard Aufderheide, Implications of eigenvector localization for dynamics on complex networks, 
PhD Dissertation, (2014).


\bibitem{hessian_matrix_loc_2009} B. J. Huang and Ten-Ming Wu, Localization-delocalization transition in Hessian matrices of topologically disordered systems, Phys. Rev. E {\bf 79}, 041105 (2009).

\bibitem{hessian_matrix_loc_2014} L. Pettinato, E. Calistri, F. D. Patti, R. Livi, and S. Luccioli, Genome-wide analysis of promoters: clustering by alignment and analysis of regular pattern, Plos One {\bf 9}, e85260 (2014).

\end{thebibliography}
\end{document}